# Hayashi Spectra of the Northern Hemisphere Mid-latitude Atmospheric Variability in the NCEP and ERA 40 Reanalyses

Alessandro dell'Aquila[(1)], Valerio Lucarini[(2)], Paolo Ruti[(1)], and Sandro Calmanti[(1)]

[1]*Progetto Speciale Clima Globale, Ente Nazionale per le Nuove Tecnologie, l'Energia e l'Ambiente, Roma, Italy*

[2]*Dipartimento di Matematica ed Informatica, Università di Camerino, Camerino (MC), Italy*

## Abstract

We compare 45 years of the reanalyses of NCEP-NCAR and ECMWF in terms of their representation of the mid-latitude winter atmospheric variability for the overlapping time frame 1957-2002. We adopt the classical approach of computing the Hayashi spectra of the 500 hPa geopotential height fields. Discrepancies are found especially in the first 15 years of the records in the high-frequency-high wavenumber propagating waves and secondly on low frequency-low wavenumber standing waves. This implies that in the first period the two datasets have a different representation of the baroclinic available energy conversion processes. In the period starting from 1973 a positive impact of the aircraft data on the Euro-Atlantic synoptic waves has been highlighted. Since in the first period the assimilated data are scarcer and of lower quality than later on, they provide a weaker constraint to the model dynamics. Therefore, the resulting discrepancies in the reanalysis products may be mainly attributed to differences in the models' behavior.

# 1. Introduction

Reanalysis products are designed to obtain global, homogeneous and self-consistent datasets of the atmospheric dynamics on the longest time scale allowed by the currently available instrumental data (Kistler et al., 2001). Recently, the National Center for Environmental Prediction (NCEP), in collaboration with the National Center for Atmospheric Research (NCAR) (Kistler et al., 2001), and the European Center for Mid-Range Weather Forecast (ECMWF) (Simmons and Gibson, 2000) have released re-analysed datasets for the time frames 1948-2004 and 1957-2002, respectively. We henceforth refer to such reanalyses with the names of NCEP and ERA 40, respectively.

The re-analysed data have been used in a large variety of contexts, ranging from the direct cross validation with independent observations (Josey, 2001; Renfew et al., 2002), to the study of specific aspects of atmospheric dynamics (Annamalai et al., 1999; Hodges et al., 2003), to their usage as external input for the modeling of other components of the climate system (Harrison et al., 2002; Josey et al., 2002). Since they represent the best available homogeneous datasets for climate studies and for validation of climate general circulation models (GCMs), it is of paramount importance to assess the consistency of these reanalyses in the representation of the variability of the atmosphere. Moreover, the reanalyses can presently be considered long enough to allow for the detection of significant changes (trends) in the climate system. Any robust change in climate occurring on the time scales of a few decades should be detectable in both reanalyses. The presence of an overlapping period of almost five decades (1957-2002) between the NCEP and ERA 40 datasets provides the baseline for intercomparison studies. In a sense, the assimilation processes employed to produce the two reanalyses are to be envisioned as slightly different dynamically consistent interpolations of similar data. In this view, the two dataset are expected to provide equivalent pictures of atmospheric variability and changes. However, the reanalyses differ under several aspects. The assimilation of data is based upon the use of slightly modified version of the numerical model employed for operational weather forecast. The numerical models employed for producing NCEP and ERA 40 differ mainly in their resolution and in the parameterization of small scale physics. The resulting discrepancies between the two datasets should be taken into account in intercomparison studies (Caires et al., 2004; Ruiz-Barradas et al., 2004; Sterl, 2004). Other inhomogeneities which are *internal* to each of the two dataset may arise from the unequal spatial and temporal density of instrumental data available during the period covered by the reanalyses (Kistler et al., 2001). Most importantly, the assimilation of satellite data has dramatically improved the quality of the reanalyses over the last three decades (Sturaro, 2003). Sterl (2004), analyzing monthly averaged variables, showed that the two reanalyses differ the most where less instrumental data are available (i.e. in the southern hemisphere). In the work of Sterl (2004), no significant difference is reported in the northern hemisphere. However, large inhomogeneities exist in the spatial and temporal distribution of data, even in the northern hemisphere. Therefore, discrepancies may exist between the two reanalyses which are not captured within the low time resolution employed by Sterl (2004).

In this work, we reconsider the differences observed in the mid-latitude northern hemispheric winters observed in the NCEP and ERA 40 reanalyses by allowing for a time resolution of one day.

We use classical space time Fourier decomposition techniques to characterize the discrepancies in the 500hPa geopotential height fields provided by the two reanalyses.

Theoretical and observational arguments suggest that two main features of mid-latitude northern hemispheric winter variability can be somewath unambiguously separated, both in terms of signal and physical processes (Benzi et al., 1986).

The synoptic phenomena are traveling waves characterized by time scales of the order of 2-7 days and by spatial scales of the order of few thousands kilometers, and can be associated with release of available energy driven by conventional baroclinic conversion (Blackmon, 1976; Speranza, 1983 Wallace et al, 1988). At lower frequencies, (period of 10-40 days) the variability is mostly due to the dynamics of long stationary waves, locked by orography (Charney and Straus, 1980; Hansen and Sutera, 1983; Buzzi et al. 1984; Benzi et al., 1986).

Building upon this conceptual framework, we first measure the difference between the two reanalyses in the spectral space explored by means of the Hayashi decomposition (Hayashi, 1971, 1979; Pratt, 1976; Fraedrich and Bottger, 1978) into standing and traveling waves. Then, we suggest some possible explanation based on actual observations of the 500hPa geopotential height field and of our knowledge of the assimilation process.

This study is also a preliminary test for some of the methodologies and diagnostics tools which will be employed in the broader context of a sub-diagnostic project presented to the Program for Climate Model Diagnosis and Intercomparison (PCMDI), sponsored by the Intergovernmental Panel on Climate Change Assessment Report 4 (IPCC-AR4).

The paper is organized as follows. In section 2 we describe the data and how spectral methods can be used to select the latitudinal band relevant for mid-latitudes atmospheric variability. In section 3 we show the results of Hayashi spectral analysis for the two datasets and discuss the emerging differences. In section 4 we compare the variability of the 500hPa geopotential height fields for the two datasets before and after the onset of satellites observations. In section 5 we present our conclusions and outlook on future studies.

## 2. Data and methods

The 500hPa geopotential height is the most relevant variable descriptive of the large scale atmospheric circulation (Holton, 1992). Therefore, it constitutes a fundamental benchmark for the comparison of different atmospheric datasets of climatological relevance. We use the freely available northern hemisphere 500hPa geopotential height reanalysis provided by the NCEP and ERA 40 datasets. We consider the datasets for the overlapping time frame ranging from September $1^{st}$ 1957 to August $31^{st}$ 2002. Both reanalyses are publicly released with spatial resolution of 2.5° x 2.5°, with a resulting effective horizontal grid of 144 x 73 points. The maximum time resolution is six hours. However, in order to focus on time scales longer than one day, for both reanalyses we consider daily data, obtained by arithmetically averaging the available 4 times daily data.

Our study focuses on the northern hemisphere mid-latitude atmospheric winter variability. To this aim, we follow Benzi and Speranza (1989) and select the December-January-February (DJF) data relative to the latitudinal belt 30°N-75°N where the bulk of the baroclinic and of the low frequency planetary waves activity is observed. We average day wise the geopotential field over such latitudinal belt in order to derive a one dimensional longitudinal field representative of the atmospheric variability in the mid-latitudes.

The variability of the one dimensional field in terms of waves of different periods and wavelengths can be effectively described by means of the space-time Fourier decomposition introduced by Hayashi (1971, 1979). By computing the cross-spectra and the coherence of the signal, such method allows for a separation of the eastward and westward wave propagating components from the standing component. In the Appendix we present a full account of this methodology.

## 3. Hayashi spectra

### 3a. Climatological average

Figures 1a-1d show the various components of the 45-winters averages of the Hayashi spectra as computed from the NCEP reanalysis dataset. The spectra express the energy density of the wave field with respect to frequency and wavenumber and its decomposition into standing and propagating components. Figure 1a shows $\overline{H}_T(k,\omega)$, the total energy spectrum, figure 1b shows $\overline{H}_S(k,\omega)$, the energy spectrum related to standing waves, figure 1c shows $\overline{H}_E(k,\omega)$, the energy spectrum related to eastward propagating waves, and figure 1d shows $\overline{H}_W(k,\omega)$, the energy spectrum of the westward propagating waves. The overbar indicates the operation of averaging over the N=45 winters. Throughout the paper, the indexes $T$, $S$, $E$, and $W$ indicate total, standing and eastward/westward propagating components, respectively. As customary in literature, the Hayashi spectra have been obtained by multiplying the energy spectra by $k \cdot \omega/2\pi$, in order to compensate for the non-constant density of points in a log-log plot (see Appendix A).

A large portion of the total energy is concentrated in the low frequency – low wavenumber domain, and can be related mostly to standing waves and to westward propagating waves, as can be deduced from figures 1b and 1d. The high frequency - small wavelength domain, corresponding mainly to synoptic disturbances, contains a smaller portion of the total energy, and is essentially related to eastward propagating waves. The fact that eastward propagating waves are essentially characterized by short wavelength and that long wavelengths characterize westward propagating waves is consistent with the Rossby waves picture. Accordingly, in figure 1c we observe that it is possible to recognize the frame of a monotonic dispersion relation $\omega = \omega(k)$ for eastward propagating waves. Instead, in figure 1d the appearance of a dispersion relation for westward propagating waves is unclear. These results have a good agreement with the past analyses performed along the same lines (Fraedrich and Bottger, 1978; Speranza, 1983).

A first insight of the discrepancies between the two reanalyses can be gained by considering the difference between the two mean Hayashi spectra, portrayed in figures 2a-2d.

When considering the difference of the two total energy spectra (figure 2a), major discrepancies can be observed in the high-frequency range, corresponding to synoptic, eastward propagating disturbances. This feature is confirmed in Fig. 2c where a significant discrepancy is observed in this component of the Hayashi decomposition. Instead, no significant difference is found in the westward propagating signal (figure 2d).

When considering the difference spectrum of the standing waves energy (figure 2b), we observe that the most significant contribution comes from the low frequency-low wavenumber sector. In particular, large differences are observed for wavenumbers ranging from 3 to 5, and for periods longer than 20 days. In this domain we do not observe large discrepancies for the total energy spectra. This implies that the two datasets should differ in the repartition of the energy between standing and traveling waves.

Such disagreements are quite impressive because the Hayashi spectra have been averaged over the 45 winters, thus proving that the climatology of the winter atmospheric variability of the two datasets is not fully equivalent.

## 3b. Interannual variability

In this paragraph we inspect the temporal behavior of the previously observed discrepancies by considering specific spectral subdomains. We introduce the following integral quantities:

$$(1) \quad E_j^n(\Omega) = \int_{\omega_1}^{\omega_2}\int_{k_1}^{k_2} d\omega dk \, H_j^n(k,\omega), \qquad \text{with } j=T,S,E,W;$$

where $n$ indicates the year. The integration extremes, $\omega_{1/2}$ and $k_{1/2}$, determine the spectral region of interest $\Omega = [\omega_1, \omega_2] \times [k_1, k_2]$. The quantity $E_j^n(\Omega)$ introduced in equation (1) represents the fraction of energy of the full spectrum associated to a given subdomain $\Omega$ and to a given year $n$. Therefore, for various choices of $\Omega$, the comparison of the quantities $E_j^n(\Omega)$ obtained from the two reanalyses can help identifying differences in the capability of describing processes occurring on a given spatial and temporal scale and pertaining to qualitatively different physical processes.

In table 1 we report the time average $\overline{E}_j(\Omega)$ for the two reanalysis, computed over the whole wave number and frequency domain. Total, standing, eastward, and westward propagating energy spectrum are reported. Following basic statistical arguments, we estimate the standard error of the time-averaged value as a function of the interannual variability of the signal: $\Delta_{E_j(\Omega)} = \frac{\sigma_{E_j(\Omega)}}{\sqrt{N}}$, where $N$ is the number of years considered in the averaging process. In the considered latitudinal band, most of the energy is due to the eastward propagating component, as can be inferred also from figure 1. In all cases the time-average of the ERA 40 signal is larger, but the discrepancies between the averaged values for NCEP and ERA 40 don't exceed the standard error $\Delta_{E_j(\Omega)}$, with the exception of $\overline{E}_E(\Omega)$ (gridded cells in the table 1). This can be considered as an indication that NCEP and ERA40 are significantly different in the description of eastward propagating features.

Inspection of the yearly differences in $E_j^n(\Omega)$ for the total, standing and eastward/westward propagating components (figure 3), reveal a systematic bias in the period 1958-1972, with a major contribution from the eastward propagating component. The mean difference in the following period is negligible.

For various choices of the spectral sub-domain $\Omega$, the comparison of the quantities $E_j^n(\Omega)$ obtained from the two reanalyses can help identifying differences in the capability of the two reanalyses in describing processes occurring on a given spatial and temporal domain and pertaining to different physical processes. In table 2 we propose a clear-cut division of the waves into four categories, on the basis of the definition of $\Omega$. The four categories are well-separated and comprehend most of the wave energy of the atmosphere. As a consequence of what shown in figures 1a-1d, we have that by computing $E_j(\Omega)$ for the spectral sub-domains defined in table 2, we obtain that most of the energy has to be attributed to the LFLW and HFHW waves. The absolute values of the total energy content for each spectral subdomain are not reported. Instead, in figure 4a-b we plot the corresponding difference time series for the sub-domains LFLW and HFHW. The other sub-domains do not show significant differences. We observe that the HFHW component (figure 4b) has a very large ratio between the bias recorded in the first 15 years and the variability of the signal. The LFLW term (figure 4a) also seems to have a bias in the first period, but of the same order of magnitude of the overall interannual variability. The systematic differences observed for the components HFHW and LFLW before 1973 correspond to discrepancies of about 8% and 3% respectively, relative to the corresponding total energy content.

# 4. Differences of the geopotential height fields before and after 1973

In section 3 we have examined the discrepancies between the two reanalyses emerging from a synthetic description of the activity of planetary and synoptic waves. We now proceed by analysing the spatial distribution of the difference in atmospheric variability reproduced by NCEP and ERA40 reanalysis.

In the previous section, a systematic bias was observed during the first 15 years of the analysis. Therefore, we now divide the whole dataset into two periods, before and after 1973. A preliminary comment is the concomitance of the date with the use of the first meteorological information from satellites and aircraft data, both assimilated in the reanalysis models (see http://www.ecmwf.int/research/era/Data_Services/section3.html).

In figures 5a-d we plot the total variability for the NCEP data for the periods 1958-72 (Fig. 5a) and 1973-2002 (Fig. 5b), and the corresponding differences with the same data for the ERA40 reanalysis (Figs. 5c-5d, respectively). The two plots of the NCEP variance show that the atmospheric variability maximizes over the oceanic basins (Blackmon, 1976). The main differences occur over these regions for the whole period covered by the reanalyses, thus suggesting that the representation of small scale features may be at the core of a systematic bias between the two reanalyses. In the Atlantic sector the two reanalysis differ strongly in the earlier period, while in the second period the accordance becomes stricter. The relative differences in the total variability for the first period over the Pacific and Atlantic regions are 10% and 3% respectively. We note also a significant discrepancy over the Okhostsk Sea, where ERA40 has more variability than NCEP (first period relative difference of the order of 10%).

The corresponding figure for the low frequency (LF) variability, obtained discarding the frequency $\omega > 2\pi/10d$ (see also table 2) in a Fourier transform of the field, is not presented, showing the same features described for figure 5.

The spatial pattern of the high frequency (HF) variability ($2\pi/7d \leq \omega \leq 2\pi/2d$, as in table 2) is reported in figure 6. The NCEP HF variance shows a characteristic feature of the synoptic disturbances, concentrated downstream of the jet maxima in storm-track regions (Blackmon et al., 1977, Speranza, 1983, Wallace et al., 1988). In the first period the ERA40 has more variability both over the Pacific and the Atlantic sectors; the relative differences are of the order of 10%, with peaks of about 20% over the Pacific on the central and exit area of the storm-track regions. The difference pattern of the Atlantic sector entails the whole region between the Hudson Bay and the Mediterranean basin. This latter difference disappears in the second period, while the Pacific discrepancy reduces to a smaller area on the eastern part of the basin. We emphasize that the discrepancies observed in the HF component cannot be explained by a corresponding change in the jet stream where the baroclinic waves grow (figure not shown).

## 5. Discussion and conclusions

The NCEP and ERA40 reanalyses have been compared for the overlapping period 1957-2002. Differences in the description of the northern hemispheric winters have been quantitatively estimated by employing the space-time Fourier decomposition introduced by Hayashi (1971, 1979). A few remarks should be made about the limits of the methodology employed in this work. By computing the cross-spectra and the coherence of the signal, the Hayashi decomposition allows for the discrimination of the propagating and standing components of the wavy patterns living on a given latitudinal circle or latitudinal belt. Therefore, Hayashi spectra can capture only fundamentally one-dimensional features. Instead, the two-dimensional dynamics of synoptic systems on the sphere have considerable impact on the evolution of features that are tangent or that cross the latitudinal belt considered for this work. The presence of such features leaves spurious traces in the spectra considered in section 3. A second limit of Hayashi spectra is that they necessarily describe features that have a periodic extension over the whole latitudinal circle. Instead, important tropospheric features in the high frequency subdomain, originate and decay at specific geographical location (Tibaldi and Molteni, 1990). Nevertheless, Hayashi spectra have proved useful in identifying some of the major biases between the two considered reanalyses.

We have focused on two main regions (HF and LF) of the total spectrum, where reasonably well distinguishable processes are thought to affect the dynamics.

Our results show that, considering the HF and LF spectral features of the atmosphere, disagreements between the two reanalyses can be found in the period prior to 1973. In particular, the HF spectral component shows a sudden improvement of the bias between the first and the second period. Instead, in the low-frequency spectral region, it is more difficult to identify a similar transition. In the high-frequency domain, the improvement is observe mainly over the Atlantic basin, while in the Pacific basin differences in the variability described by the two reanalyses remain approximately constant before and after 1973.

The coverage of the Northern Hemisphere by radiosondes, dominant over the land regions, is relatively good and, to a large extent, uniform throughout the period 1958-1973 and further (Uppala et al., 2004). Therefore, the main discrepancies between the two reanalysis in the early period should be due to other data source..

The analysis performed by Bengtsson et al. (2004) highlights the positive impact of the radiosondes data over the land and of the satellite data over the ocean. Our work confirms this behavior in the reanalysis and suggests also a relevant positive impact of the aircraft data, introduced in the data assimilation process in 1973. In fact the main improvement in the HF component of the Hayashi spectra can be geographically located over the Atlantic Ocean (see figure 6), where the aircraft data are mainly available. Besides, over the Pacific Ocean basin, where the additional data are mainly from satellites, a weaker improvement is observed in the atmospheric baroclinic waves description . These results suggest a caveat in using the reanalysis data for the analysis of the low and high frequency variability and trends prior to 1973. The investigation performed in this work provides us some confidence level to use the reanalyses to validate the 500 hPa geopotential height variability in climate models.

Thereby, our results agree with the findings of Hodges et al. (2003) and Bengtsson et al. (2004) that when considering the analyzed fields prior to 1979 caution should be used because the analyses can be influenced by the model deficiencies.

It remains to understand why the two models have different behaviors, both in the HF and LF components , in the first period where data available are scarcer. The possibility of verifying the actual relevance of our interpretation is out of our reach since we can not re-run portions of the assimilation processes for both reanalyses and check for consistency. However, we suggest some possible causes of the observed discrepancies.

Dynamically, the simultaneous occurrence of significant discrepancies in both the HF and LF spectral sub-domains may have two fundamentally different interpretations. If the two spectral subdomains were dynamically independent, then the two reanalysis procedures should differ both in the treatment of physical processes which are responsible for LF variability and for processes related to the development of synoptic disturbances. However, robust arguments about energy cascades in geostrophic turbulence, suggest that energy can be transferred from high to low-frequency. In this case, a difference observed in the standing component could be attributed to the transfer of energy from smaller to larger scale. (Savijarvi, 1978; Weeks et al., 1997). In support to the latter argument, we found that a strong bias in the HFHW sub-domain exist in the first period and corresponds to a noticeable bias in the LFLW.

After 1973, when the bias in the HFHW disappears, the bias in the LFLW also disappears. However, the variability of the interannual discrepancies in the LFLW sub-domain, remains larger than the initial bias itself, indicating that a mechanism during all the reanalysis period acts to produce this error.

Such discrepancies in the description of basic physical processes may be attributed most likely to one or both of the following characteristics of the models used for the reanalyses.

The first is the resolution of the two models. The two models used for ERA40 and NCEP reanalyses use spectral methods in the horizontal and finite difference methods in the vertical. They mainly differ in the level of the truncation in the horizontal (T63 for NCEP, T159 for ERA40) and in the number of vertical levels (28 for NCEP, 60 for ERA40). The truncation level can influence the enstrophy inverse cascade, and it can also affect the horizontal convergence of momentum, which in turn alters the horizontal wind shear and controls the growth of the baroclinically unstable eddies. Moreover, the vertical resolution should impact the meridional and vertical heat fluxes associated to baroclinic perturbations that modify the thermal vertical structure of the atmosphere in the midlatitudes (Barry et al., 2000; Dell'Aquila, 2004).

Another aspect of the reanalyses ,which has been demonstrated to affect the low frequency variability ,is the envelope orography, i.e. subgrid-scale orography parameterization (Wallace et al., 1983; Tibaldi 1986). ERA40 uses an envelope orography, whereas NCEP has a mean orography. For this parameterization the smaller scales of the orography directly affect the zonal flow (Tibaldi, 1986). This fact could explain the variability of the interannual discrepancies in the LFLW sub-domain in the period after 1973.

**Acknowledgments.**

The authors wish to thank A. Sutera and A. Speranza for useful suggestions. NCEP data have been provided by the NOAA-CIRES Climate Diagnostics Center, Boulder, Colorado, from their web site at http://www.cdc.noaa.gov/. The ECMWF ERA-40 data have been obtained from the ECMWF data server at http://data.ecmwf.int/data/.

## Appendix: Space-time spectral analysis

The space-time spectral analysis, introduced by Hayashi (1971), provides information about the direction or speed at which the eddies move.

This information may be obtained by, firstly, Fourier-analysing the spatial field and then computing the time-power spectrum of each spatial Fourier component. The difficulty here lies in the fact that straightforward space-time decomposition will not distinguish between standing and traveling waves: a standing wave will give two spectral peaks corresponding to traveling waves moving eastward and westward at the same speed and the same phase. The problem can only be circumvented by making assumptions regarding the nature of the wave. For instance, we may assume complete coherence between the eastward and westward components of standing waves and attribute the incoherent part of the spectrum to real traveling waves (Pratt, 1976, Fraedrich and Bottger, 1978; Hayashi,1979).

In this formulation, for each winter considered, the energy spectrum $H_{E/W}(k,\omega)$ at a zonal wavenumber $k$ and temporal frequency $\omega$ for the eastward and westward propagating waves is :

(A1a) $\quad H_E(k,\omega) = \frac{1}{4}\{P_\omega(C_k) + P_\omega(S_k)\} + \frac{1}{2}Q_\omega(C_k, S_k)$

(A1b) $\quad H_W(k,\omega) = \frac{1}{4}\{P_\omega(C_k) + P_\omega(S_k)\} - \frac{1}{2}Q_\omega(C_k, S_k)$

$P_\omega$ and $Q_\omega$ are, respectively the power and the quadrature spectrum of the longitude ($\lambda$) and the time ($t$) dependent 500hPa geopotential height $Z(\lambda,t)$ expressed in terms of the zonal Fourier harmonics:

(A2) $\quad Z(\lambda,t) = Z_0(\lambda,t) + \sum_{1}^{\infty}\{C_k(t)\cos(k\lambda) + S_k(t)\sin(k\lambda)\}.$

The total variance spectrum $H_T(k,\omega)$ is given from the sum of the eastward and westward propagating components:

(A3) $\quad H_T(k,\omega) = \tfrac{1}{2}(P_\omega(C_k) + P_\omega(S_k))$

while the propagating variance $H_P(k,\omega)$ is given by the difference between the components (A1a) and (A1b):

(A4) $\quad H_P(k,\omega) = |Q(k,\omega)|.$

So, the standing variance spectrum $H_S(k,\omega)$ can be obtained by the difference:

(A5) $\quad H_S(k,\omega) = H_T(k,\omega) - |Q(k,\omega)|.$

We emphasize that for sake of simplicity of the notation we have neglected the indication of the winter under investigation, denoted in the text by the superscript $n$.

We emphasize that customarily, Hayashi spectra are generally represented by plotting the quantities $k \cdot \omega/2\pi \cdot H_T(k,\omega)$, $k \cdot \omega/2\pi \cdot H_S(k,\omega)$, $k \cdot \omega/2\pi \cdot H_E(k,\omega)$, and $k \cdot \omega/2\pi \cdot H_W(k,\omega)$, as reported in figures 1a-1d, in order to allow for equal geometrical areas in the log-log plot representing equal variance.

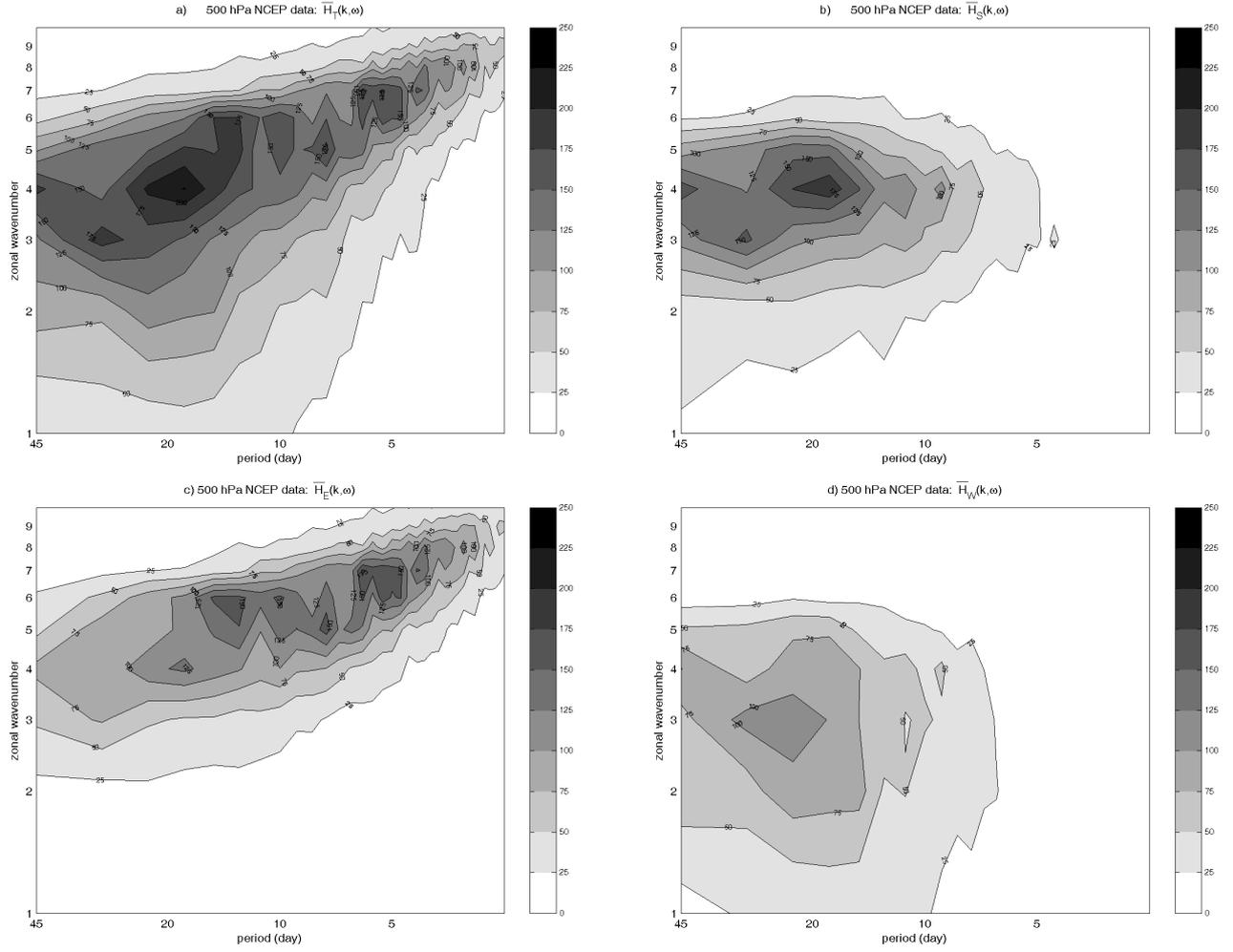

Figure 1: Climatological average over 45 winters of Hayashi spectra for 500 hPa geopotential height (relative to the latitudinal belt 30°N-75°N) from NCEP data: $\overline{H}_T(k,\omega)$ (a); $\overline{H}_S(k,\omega)$ (b); $\overline{H}_E(k,\omega)$ (c); $\overline{H}_E(k,\omega)$ (d). The Hayashi spectra have been obtained multiplying the energy spectra by $k \cdot \omega / 2\pi$. The units are *m²/s * 10⁻⁵*

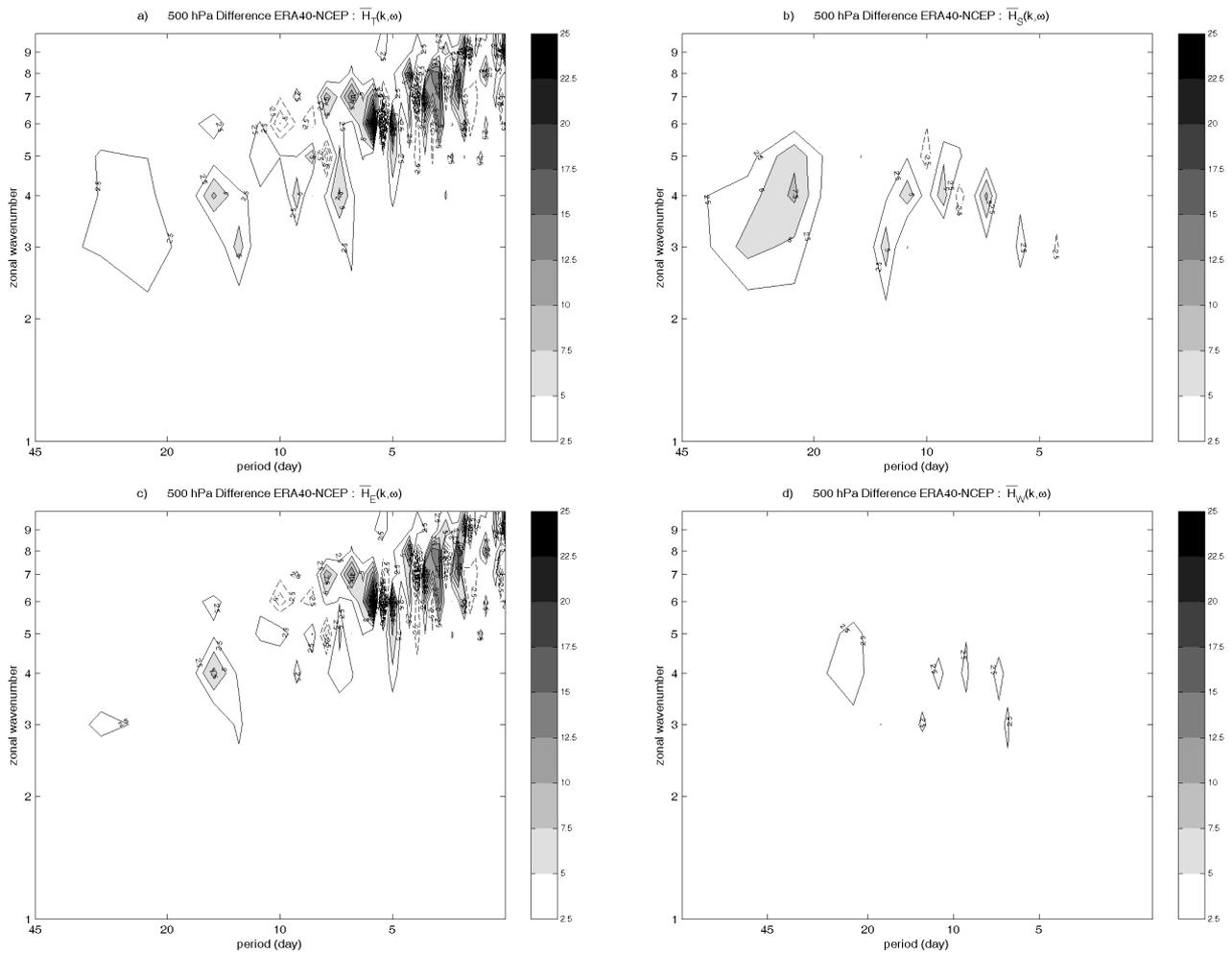

Figure 2: As for figure 1 but for the difference between ERA40 and NCEP dataset Hayashi spectra. Dashed lines are pertaining to negative values.

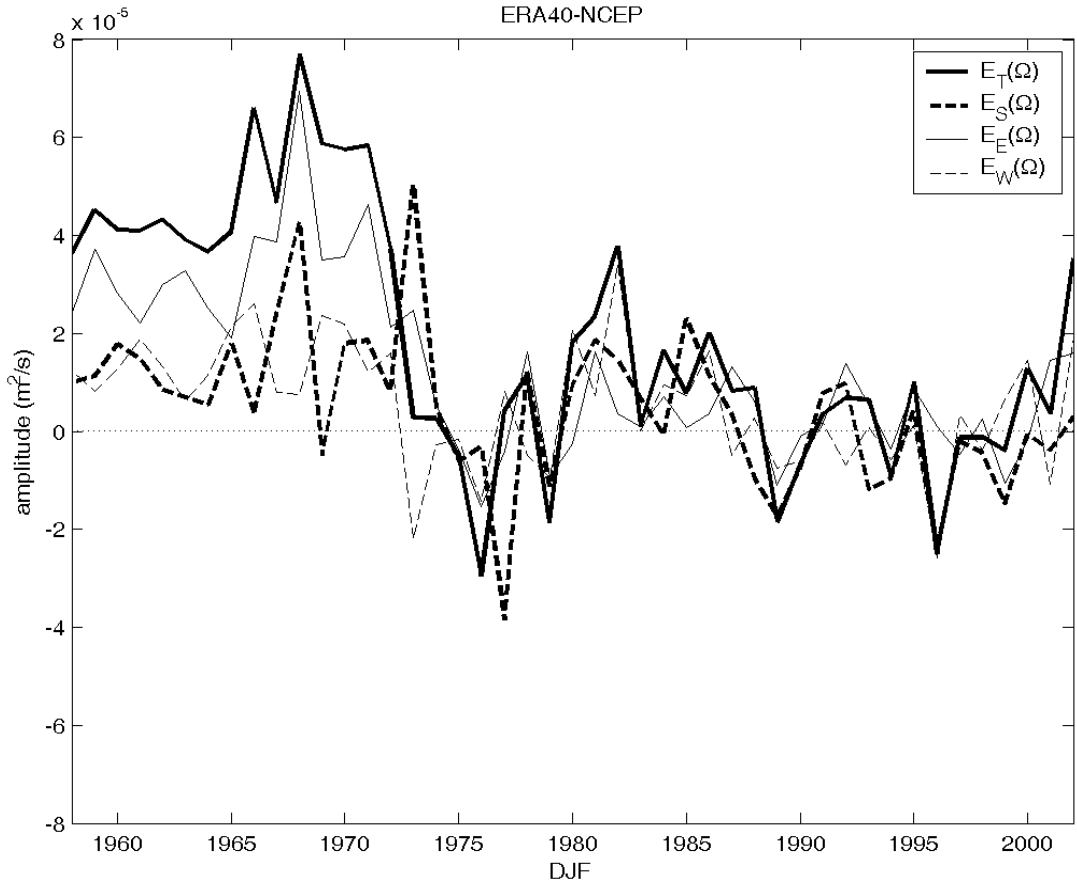

Figure 3: Time series of the differences of the quantity $E_j^n(\Omega)$ computed for the two reanalysis datasets, for: total variance (thick line), standing (dashed thick line), propagating eastward (thin line) and propagating westward (dashed thin line). Here, $\Omega$ corresponds to the whole wave number and frequency domain.

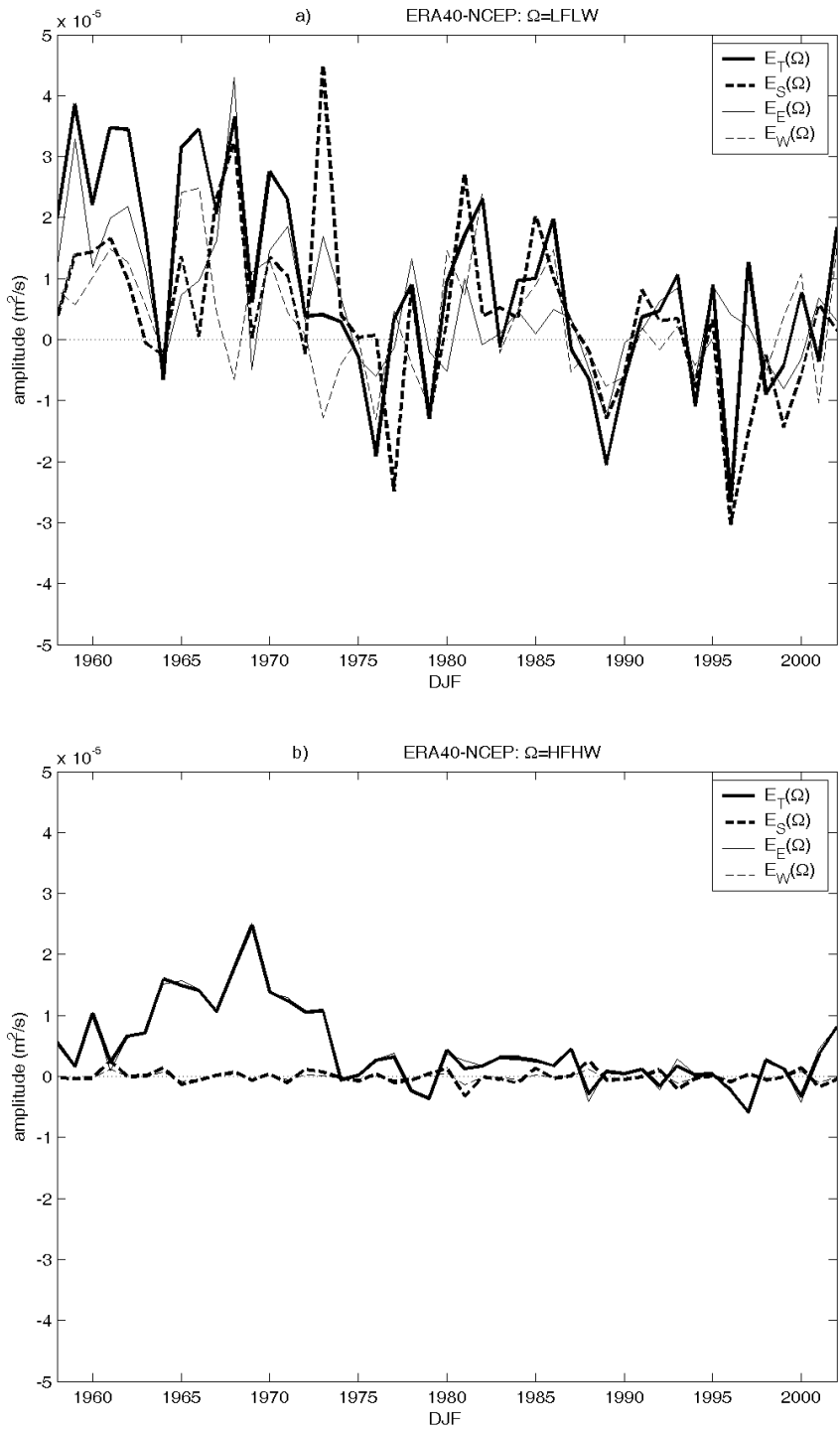

Figure 4: As in figure 3 for the categories *LFLW (a), HFHW (b)* described in table 2.

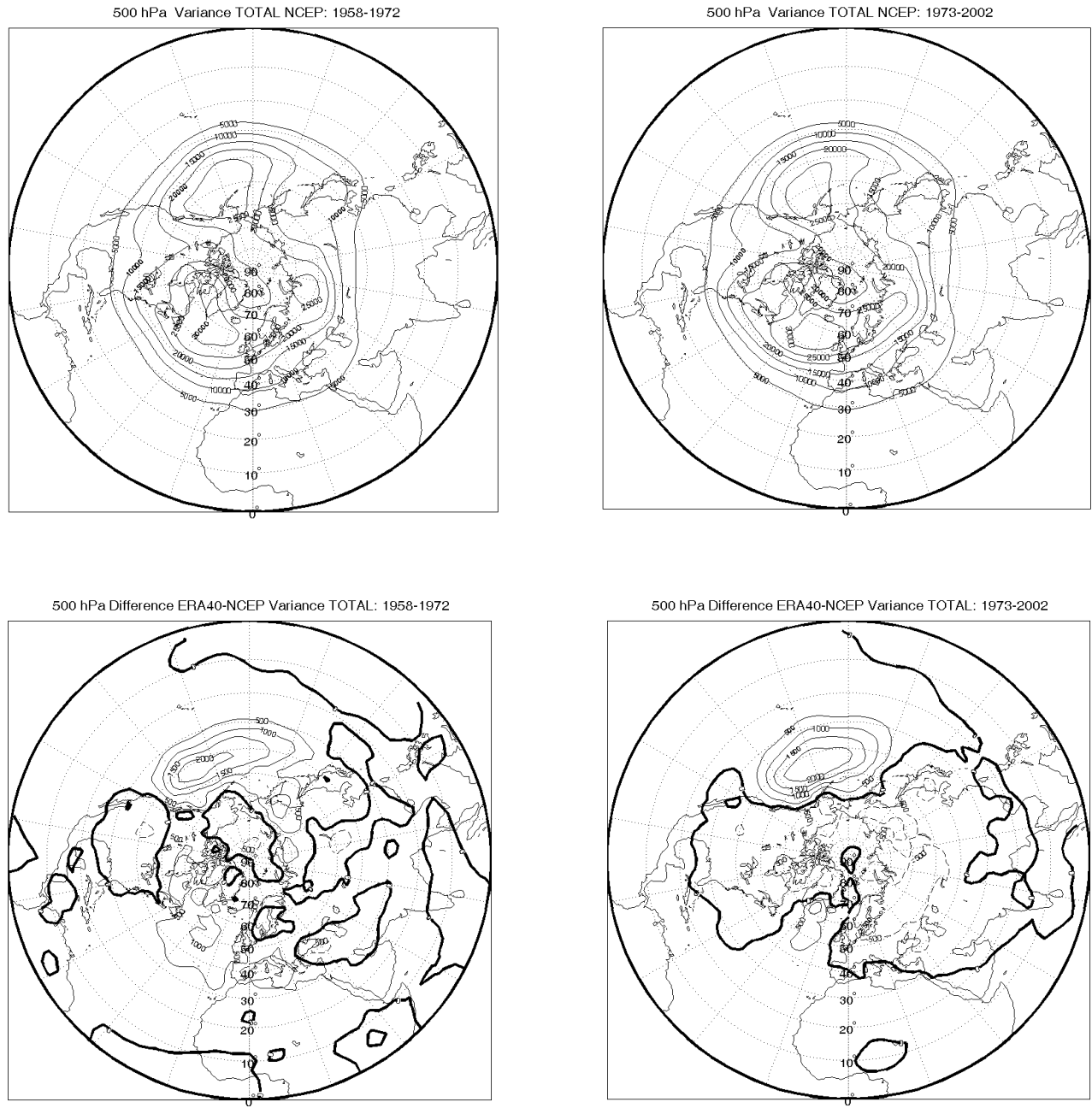

Figure 5: Total variance of the 500 hPa geopotential height, DJF period: a) NCEP 1958-1972; b) NCEP 1973-2002; c) difference between ERA40 and NCEP, 1958-72; d) as c) for 1973-2002.

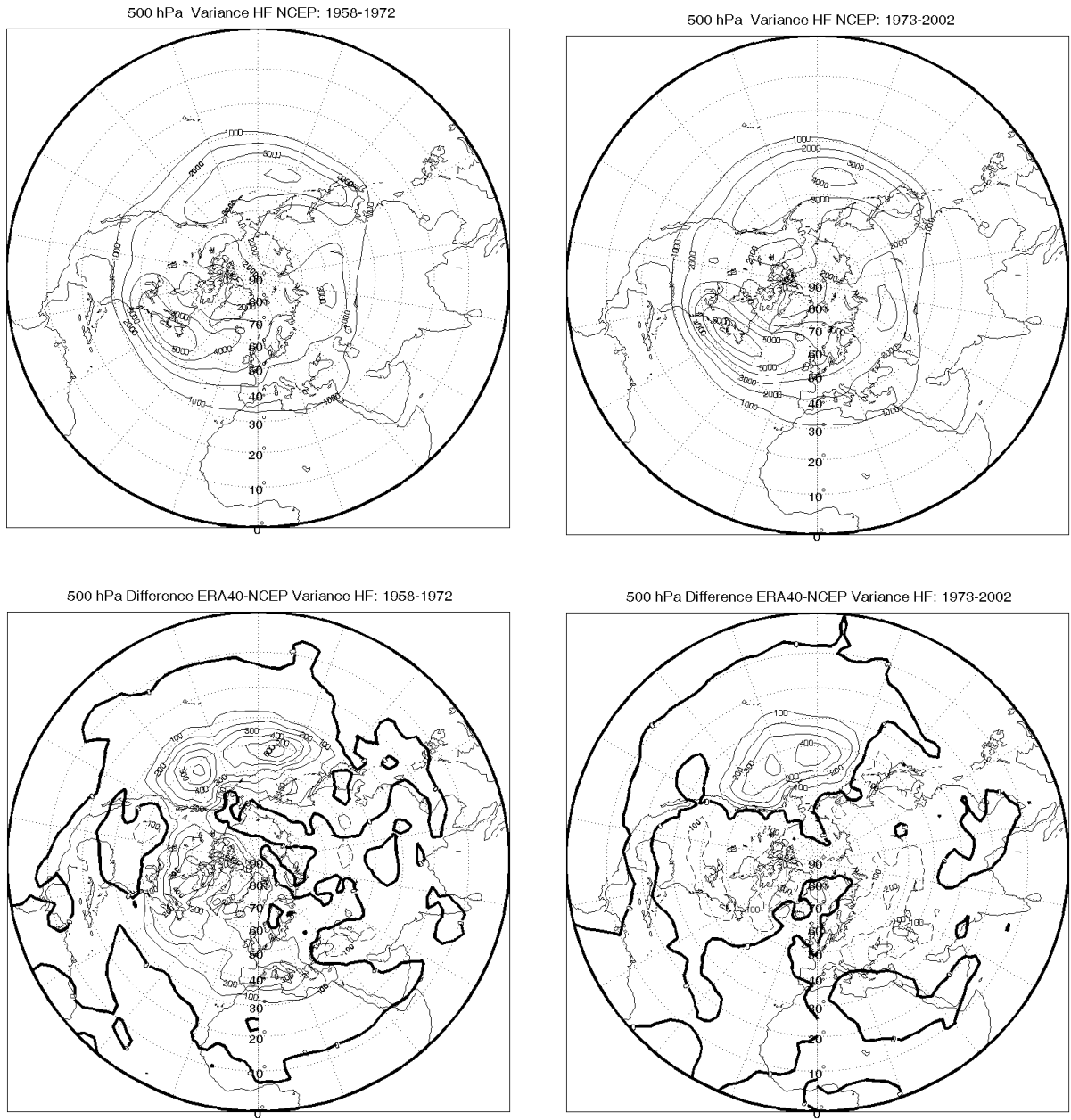

Figure 6 High frequency variance of the 500 hPa geopotential height, DJF period: a) NCEP 1958-1972; b) NCEP 1973-2002; c) difference between ERA40 and NCEP, 1958-72; d) as c) for 1973-2002.

TABLES

| $\Omega$=ALL | ERA 1957-2002 | NCEP 1957-2002 | ERA 40 1957-72 | NCEP 1957-72 | ERA 40 1973-2002 | NCEP 1973-2002 |
|---|---|---|---|---|---|---|
| $\overline{E}_T(\Omega)$ | 174 ±3 | 173±3 | 174±6 | 169±6 | 175±3 | 174±3 |
| $\overline{E}_S(\Omega)$ | 66±2 | 65±2 | 65±3 | 64±2 | 66±2 | 66±2 |
| $\overline{E}_E(\Omega)$ | 92±1 | 91±1 | 90±2 | 86±2 | 94±2 | 93±2 |
| $\overline{E}_W(\Omega)$ | 82±3 | 81±3 | 84±5 | 83±5 | 81±3 | 81±3 |

Table 1: Time mean of $E_j(\Omega)$ with $j=T, S, E, W$ for the DJF period of the whole record, 1957-1972 and 1973-2002, respectively. We consider the standard error of the time-averaged value as a function of the interannual variability of the signal: $\Delta_{E_j(\Omega)} = \frac{s_{E_j(\Omega)}}{\sqrt{N}}$. As in Fig.3, $\Omega$ corresponds to the whole wave number and frequency domain. The units are $m^2/s * 10^{-5}$

| Spectral properties | $k_1 = 2, k_2 = 4$ | $k_1 = 6, k_2 = 72$ |
|---|---|---|
| $w_1 = 2p/45d$, $w_2 = 2p/10d$ | $\Omega$ = LFLW | $\Omega$ = LFHW |
| $w_1 = 2p/7d$, $w_2 = 2p/2d$ | $\Omega$ = HFLW | $\Omega$ = HFHW |

Table 2: Definition of 4 regions in the Hayashi spectra of the winter atmospheric variability. *LFLW*: Low Frequency Long Wavenumber; *LFHW*: Low Frequency High Wavenumber; *HFLW*: High Frequency Long Wavenumber; *HFHW*: High Frequency High Wavenumber. The values $w_2 = 2p/2d$, $k_2 = 72$ constitute the highest frequency and wavenumber allowed by the adopted data resolution.